# CAN RECENTLY DERIVED SOLAR PHOTOSPHERIC ABUNDANCES BE CONSISTENT WITH HELIOSEISMOLOGY?

Joyce Ann Guzik[1], L. Scott Watson[1,2]

[1]Los Alamos National Laboratory, X-2, MS T085, Los Alamos, NM 87545 USA, Email: joy@lanl.gov
[2]Dept. of Physics and Astronomy, U. New Mexico, 800 Yale Blvd. NE, Albuquerque, NM 87131, USA, Email: swatson@lanl.gov, Scott.Watson@USAFA.af.mil

**ABSTRACT**

Recent solar abundance analyses (Asplund et al. 2004; Lodders 2003) revise downward the abundances of C, N, O, Ne, and Ar, which reduces the solar photospheric Z/X to 0.017, and Z to ~0.013. Solar models evolved with standard opacities and diffusion treatment using these new abundances give poor agreement with helioseismic inferences for sound speed profile, convection zone helium abundance, and convection zone depth. Here we present helioseismic results for evolved solar models with these reduced photospheric abundances, trying varying diffusion treatments. We compare results for models with no diffusion, enhanced thermal diffusion, and enhanced diffusion of C, N, O, Ne, and Mg only. We find that while each of these models provides some improvements compared to a solar model evolved with the new abundances and standard physics, none restores the good agreement with helioseismology attained using the earlier abundances of, e.g., Grevesse & Sauval (1998). We suggest that opacity increases of about 20% for conditions below the convection zone, or the possibility of accretion of lower-Z material at the surface as the sun arrived at the main sequence, should be investigated to restore agreement. In addition, the new abundance determinations should be reconsidered, as, if they are correct, it will be difficult to reconcile solar models with helioseismic results.

## 1. INTRODUCTION

Recent solar photospheric abundance analyses (Asplund et al. 2004; Lodders 2003) revise downward the abundances of C, N, O, Ne, and Ar, which reduces the solar photospheric Z/X to 0.017 (Lodders 0.0177) and Z to ~0.0126 (Lodders 0.0133).

Standard solar models that give good agreement with helioseismology have been calibrated to earlier higher abundance determinations, e.g., the Grevesse & Sauval (1998) values of Z/X = 0.023 and Z ~0.0171.

Basu & Antia (2004) and Bahcall & Pinsonneault (2004) concluded that solar models evolved with the new lower abundances give worse agreement with helioseismically-inferred sound speed profile, convection zone base radius, and convection zone helium abundance.

Asplund et al. suggest that enhanced diffusion may be able to restore the agreement with convection zone depth, but Basu & Antia (2004) find that it would also decrease the convection zone Y abundance below the seismically-determined value.

Here we present results for solar models evolved with different initial abundances and diffusion treatments than previously published to see if we can reconcile the new abundances and helioseismology.

Note that there are several other preprints and posters given at this meeting that explore additional ideas, such as enhanced opacities, and diffusion treatment options (e.g., posters by Montalban et al., Serenelli et al., and Basu & Antia, and astro-ph preprints by Bahcall et al. and Turck-Chieze et al. 2004).

## 2. SOLAR MODEL PROPERTIES

For our solar model and oscillation frequency calculations, we have used the codes and procedures described in Neuforge-Verheecke et al. (2001a,b) and references therein. We use the LLNL OPAL (Iglesias & Rogers 1996) opacities and the Alexander and Ferguson (1995) low-temperature opacities, both with the Grevesse & Noels (1993, GN93) solar mixture. We have calibrated our models to a Z/X somewhat larger than that recommended by Asplund et al., since the GN93 mixture contains relatively less Fe than the Asplund et al. mixture, and so we can roughly compensate for the mixture differences by using a slightly higher Z. We use the SIREFF analytical equation of state (Guzik & Swenson 1997) to account for the changes in element mixtures. However, we find that accounting for the relatively small mixture changes between GN93 and Asplund et al. in the EOS has a very small effect on the model structure compared to the overall decrease in Z we are investigating. We use the NACRE (Angulo et al. 1999) nuclear reaction rates and standard mixing-length convection treatment (Bohm-Vitense 1958).

We use the Burgers (1969) diffusive element settling treatment as implemented by Cox, Guzik & Kidman (1989) that includes thermal, gravitational, and chemical diffusion of H, He, C, N, O, Ne, and Mg. Since the elements are treated individually via separate coupled equations, we can experiment with adjusting the resistance coefficients for individual elements, which are uncertain in any case, to allow enhanced diffusion of C, N, O, Ne, and Mg while avoiding the diffusion of too much helium.

The models are calibrated to the present solar radius ($6.9599 \times 10^{10}$ cm), luminosity ($3.846 \times 10^{33}$ erg/s), mass ($1.9891 \times 10^{33}$ g), and age ($4.52 \pm 0.04$ Gyr; Guenther et al. 1992).

For reference, we also quote here the constraints from helioseismic inversions of Basu & Antia (2004) on the convection zone Y ($0.248 \pm 0.003$), and convection zone base radius ($0.7133 \pm 0.0005$ $R_{sun}$).

## 3. SOLAR MODEL COMPARISONS

We compare five evolved models: 1) A standard solar model with GN93 abundances and standard diffusion treatment; 2) a low-Z model with reduced C, N, O, Ne, and Mg abundances consistent with the Asplund et al. (2004) abundances, and no diffusion; 3) a model with the same initial Z as Model 1, but with the thermal diffusion enhanced for C, N, O, Ne, and Mg only to enhance selectively their diffusion and avoid too much helium diffusion. For this model, we lowered the binary thermal resistance coefficients for the motion of these elements relative to hydrogen by a factor of 7; 4) a model with the same initial Z as the standard model, but with the binary thermal resistance coefficients lowered for all elements by a factor of 3; 5) a model with high initial Z (0.025) that consequently also has high initial Y (0.284), and binary thermal resistance coefficients for C, N, O, Ne, and Mg reduced by a factor of 15.

Table 1 compares the initial abundances, final surface abundances, and final properties of the first four evolved models for which we completed the seismic analysis (Figs. 1 and 2). Fig. 1 compares the sound speed profile difference for each model with the sound speed inversion results of Basu et al. (2000). Fig. 2 compares the observed minus calculated frequency differences for each model for modes of degree $l$=0, 2, 10, and 20. The frequency calculations are done using the nonadiabatic code of Pesnell (1990). The observed frequencies are from the BiSON group (Chaplin et al. 1998) the LowL group (Schou & Tomczyk 1996).

Table 1. Properties of evolved models

| | Standard GN93 | Low-Z No Diffusion | Enhanced Z Diffusion | Enhanced Diffusion |
|---|---|---|---|---|
| $Y_o$ | 0.2703 | 0.2493 | 0.2650 | 0.2626 |
| $Z_o$ | 0.0197 | 0.01425 | 0.0197 | 0.0197 |
| $Y_{surf}$ | 0.2419 | 0.2493 | 0.2339 | 0.1926 |
| $Z_{surf}$ | 0.01805 | 0.01425 | 0.01400 | 0.01552 |
| Z/X | 0.0244 | 0.0194 | 0.0186 | 0.0196 |
| $\alpha$ | 1.769 | 1.560 | 1.658 | 1.944 |
| $R_{czb}/R_{sun}$ | 0.7133 | 0.7388 | 0.7283 | 0.7022 |
| $T_c$ $10^6$ K | 15.66 | 15.21 | 15.69 | 15.83 |
| $^{37}$Cl vs SNUs | 7.78 | 4.80 | 7.90 | 9.12 |

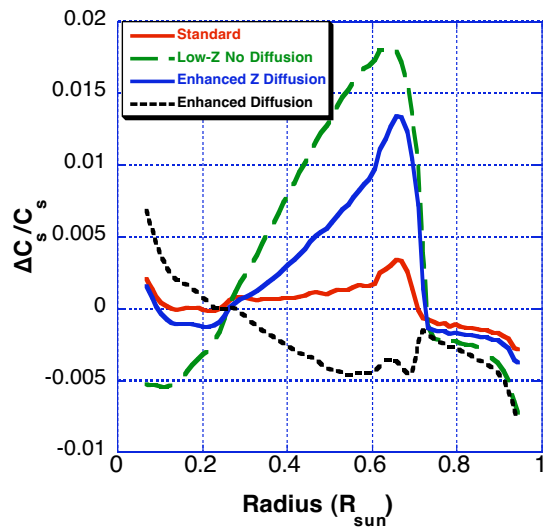

Figure 1. Sound speed profile differences [(seismic-model)/seismic] for four models. Seismic inversion from Basu et al. (2000).

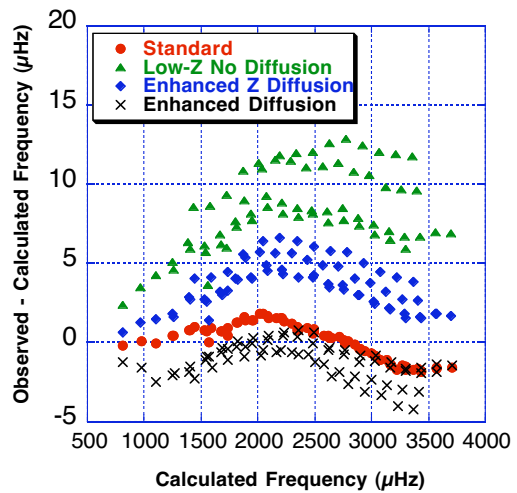

Figure 2. Observed minus calculated frequencies versus frequency for degrees l=0, 2, 10, and 20. Observations are from either BiSON (Chaplin et al. 1998) or LowL (Schou & Tomczyk 1996).

## 4. MODEL EVALUATION/DISCUSSION

One can see from Table 1 and Figs. 1 and 2 that the agreement with helioseismology for the standard model with the GN93 abundances and standard diffusion treatment and opacities is excellent, although small improvements are still needed.

The no-diffusion reduced-Z model has the advantage that the lower Z also requires a lower initial Y (0.249) to match the present solar luminosity, so that the initial (and therefore present convection zone) Y agrees with helioseismology without the need for including any diffusion! However, this model gives poor results for the sound speed profile below the convection zone, with the discrepancies as large as 18%, compared to only 4% for the standard model. The convection zone of this model is also very shallow (base radius 0.7388 $R_{sun}$).

For Model 3 with enhanced diffusion of selected elements only, the convection zone is still somewhat too shallow (base radius 0.729), and the surface Y is slightly low (0.234). These discrepancies might be reduced with modest opacity increases below the convection zone. However, there is no justification for the large *ad hoc* lowering of the resistance coefficients for selected elements.

For Model 4 with *ad hoc* lowering of the binary resistance coefficients for all elements relative to hydrogen, the final surface Y is much too low (0.1926), while the surface Z is not low enough (0.0155). This model has a convection zone that is too deep, and the sound speed discrepancies are about 5% both below the convection zone and near the solar center. Adjustments to the initial Z and diffusion rates that would remedy the problems with the convection zone base and final Z will still result in a convection zone Y that is too low. Basu & Antia (2004) also found this problem of a too-low convection zone Y for enhanced-diffusion models.

We also attempted to improve the sound speed agreement below the convection zone by increasing the initial Z to 0.025, and greatly enhancing the diffusion of C, N, O, Ne, and Mg by lowering the binary thermal resistance coefficients by a factor of 15 (Model 5). For this model, while the final convection zone Y remains high enough (Y=0.252), the convection zone depth is too shallow (base radius 0.732 $R_{sun}$). This model has final surface Z=0.0140, and Z/X = 0.0191, in fair agreement with the new abundances.

## 5. CONCLUSIONS AND FUTURE WORK

As discussed by Basu & Antia (2004), and Bahcall & Pinsonneault (2004), the new photospheric element abundances give worse agreement with helioseismology. The agreement can be improved somewhat by enhanced diffusion of elements C, N ,O, Ne, & Ar relative to H, He, but not enough to restore agreement attained with the standard model.

We are forced to question whether something has been overlooked in the revised abundance determinations of Asplund et al. that is causing them to be systematically too low. However, their papers are very convincing given the consistency in the new abundances from several indicators, and the many improvements incorporated into the analysis.

Judging from comparisons of solar models with the OPAL and slightly lower LANL LEDCOP opacities (Neuforge-Verheecke et al. 2001b), opacity increases of about 20% above the OPAL values for conditions below the convection zone would nearly eliminate the discrepancies in sound speed and convection zone depth. Bahcall et al. (2004a) also suggest that opacity increases of this magnitude are needed just below the convection zone to deepen the convection zone to the seismically-determined value, and Bahcall et al. (2004b), find that opacity increases of 11% between 2 and 5 million K are needed to improve the sound speed profile agreement. However, the LLNL OPAL and LANL LEDCOP opacities, calculated independently with different approaches, now agree for solar conditions to within a few percent (neglecting differences due to coarse temperature and density grid spacing and interpolation). It may be unlikely, considering this agreement, that the Rosseland mean opacities for solar mixtures could be incorrect by more than several percent for conditions below the convection zone. The importance of resolving this discrepancy with helioseismology provides good motivation for opacity experiments at laser or pulsed-power facilities, as suggested by Turck-Chieze et al. 2004). Such a large discrepancy with theory may be detectable experimentally, whereas a few percent difference would not be measurable, due to intrinsic experimental uncertainties.

Several groups (see, e.g., posters by Serenelli et al. and Montalban et al.) are considering whether agreement can be restored by adopting some combination of effects, such as enhanced diffusion plus opacity increases, or adopting the highest abundances allowed by the uncertainties, combined with opacity increases, also within limits allowed by uncertainties. Bahcall et al. (2004b) point out that the Opacity Project (OP) opacities are 5% higher than the OPAL opacities for conditions below the convection zone for a six-element mixture with Z=0.02 (Seaton & Badnell 2004).

We also suggest considering the remote possibility of mass *accretion*. Perhaps a model could be evolved that is consistent with helioseismology if the initial ~98% of the sun's mass accumulated during its formation had higher Z, coincidentally nearer that of the GN93 mixture. The last ~2% of material accreted would need to have somewhat lower Z, consistent with the present abundances accounting for an expected amount of diffusion. The accretion would need to occur after the sun arrives on the main sequence and is no longer fully convective, but it could occur very early, within a few million years of the sun's arrival on the main sequence. However, this solution would pose its own problems for solar evolution, as then the conditions of the sun's formation are not well-constrained, and more parameters are introduced. If the only way out of the present dilemma is to postulate very different abundances in the solar interior from those detectable spectroscopically, we may never be able to infer unambiguously the sun's interior abundance and decouple it from opacity, diffusion treatment, or EOS uncertainties.